\documentclass[]{raa}            
     \usepackage{longtable}
\usepackage{graphicx,times}             

\begin{document}

   \title{The comparison of H$_2$CO (1$_{10}$--1$_{11}$), C$^{18}$O (1--0) and continuum towards molecular clouds
}

   \volnopage{Vol.0 (2014) No.0, 000--000}      
   \setcounter{page}{1}          

   \author{X. D. Tang
      \inst{1,2}
   \and J. Esimbek
      \inst{1,3}
   \and J. J. Zhou
      \inst{1,3}
      \and G. Wu
      \inst{1,2,3}
   \and D. Okoh
      \inst{1,4}
   }

   \institute{Xinjiang Astronomical Observatory, Chinese Academy of Sciences, Urumqi 830011, China {\it tangxindi@xao.ac.cn}\\
        \and
             University of the Chinese Academy of Sciences, Beijing 100080, China\\
        \and
             Key Laboratory of Radio Astronomy, Chinese Academy of Sciences, Urumqi 830011, China\\
        \and
             Physics \& Astronomy Department, University of Nigeria, Nsukka 410001, Nigeria\\
   }

\date{Received~~2009 month day; accepted~~2009~~month day}

\abstract{We present large scale observations of C$^{18}$O (1--0) towards four massive star forming regions for MON R2, S156, DR17/L906 and M17/M18. The transitions of H$_2$CO (1$_{10}$--1$_{11}$), C$^{18}$O (1--0) and 6 cm continuum were compared towards the four regions. Analysis of observation and Non--LTE model results shows that the brightness temperature of the formaldehyde absorption line is strongest in background continuum temperature range of about 3 -- 8 K. The excitation of the H$_2$CO absorption line is affected by strong background continuum emission. From the comparison of H$_2$CO and C$^{18}$O maps, we found that the extent of H$_2$CO absorption is broader than that of C$^{18}$O emission in the four regions. Except for the DR17 region, the H$_2$CO absorption maximum is located at the same position with the C$^{18}$O peak. The good correlation between intensities and widths of H$_2$CO absorption and C$^{18}$O emission lines indicate that the H$_2$CO absorption line can trace dense and warm regions of the molecular cloud. Finding that N(H$_2$CO) was well correlated with N(C$^{18}$O) in the four regions and that the average column density ratio is $<$N(H$_2$CO)/N(C$^{18}$O)$>$ $\sim$ 0.03.
\keywords{ISM: clouds -- molecules -- stars: formation}
}

   \authorrunning{X. D. Tang, J. Esimbek, J. J. Zhou, G. Wu \& D. Okoh }            
   \titlerunning{The comparison of H$_2$CO, C$^{18}$O, and continuum towards molecular clouds}  

   \maketitle

\section{Introduction}           
\label{sect:intro}

Dense molecular cloud cores are considered to be sites of star formation. Studying the structure and condition of these cores allows us to have a better understanding of the whole process of star formation. It is important to have a comprehensive understanding of the evolution of dense regions.

The first detection of H$_2$CO was reported by Snyder et al. (1969). It is the first organic polyatomic molecule discovered in the interstellar medium. The dense clumps in the line of sight may be the region of H$_2$CO production (Prasad \& Huntress 1980). The formaldehyde absorption provides a powerful tool for analyzing the distribution of dense interstellar gas (Scoville \& Solomon 1972). For the typical densities of clouds near HII regions (10$^{3}$ to 10$^{5}$ cm$^{-3}$) the 1$_{10}$--1$_{11}$ transition of H$_2$CO is ideal (Gardner et al. 1984). The 4.83 GHz 1$_{10}$--1$_{11}$ transition of interstellar H$_2$CO is easily observed in absorption against galactic continuum source. And it is also easily seen in absorption toward many dark clouds where the molecule's  lines are absorbing the cosmic background radiation (CMB).

Downes et al. (1980) found that the intensity of the H$_2$CO absorption line is affected strongly by the background continuum emission in the HII region. Zhang et al. (2012) obtained a similar result in giant molecular clouds, and inferred that the contribution of CMB radiation to the H$_2$CO intensity was very weak. The giant HII regions have many compact continuum components, and the H$_2$CO absorption is affected by the distribution of these continuum components. A mapping study of 12 bright galactic HII regions and 2 dark clouds showed that the strong H$_2$CO absorption were quite away from the bright continuum peaks in a number of cases (Bieging et al. 1982). Pipenbrink \& Wendker (1988) surveyed the brightest features in the Cyg X region and found that there are 22 positions where no H$_2$CO absorption line was detected. This is especially surprising in the high brightness levels of HII regions.

Recently, Tang et al. (2013) studied the correlation among H$_2$CO (1$_{10}$--1$_{11}$), $^{12}$CO (1--0), and $^{13}$CO (1--0) towards the MON R2, S156, DR17/L906 and M17/M18 regions, and they found that the H$_2$CO and $^{13}$CO integral intensity maps had analogous shapes, sizes, peak positions and molecular spectra presenting similar central velocities and line widths. Such a good agreement indicates that the H$_2$CO absorption and the $^{13}$CO emission lines arise from similar regions. The $^{12}$CO (1--0) line was not suitable for tracing the star formation region because of the larger optical depth. The $^{13}$CO (1--0) spectra tend to be saturated towards the high density regions, making it difficult to understand the detailed distribution of the high density regions. The abundance of C$^{18}$O molecule is less than that of $^{13}$CO molecule by a factor of $\sim$ 5. The C$^{18}$O line is optically thin compared to $^{13}$CO and could trace dense regions ($>$ 10$^{4}$ cm$^{-3}$) (Dame et al. 1986). Therefore, the C$^{18}$O is a good spectral line to trace dense regions in nearby star formation regions. Minn et al. (1996) had compared the distributions of the H$_2$CO 6-cm absorption and C$^{18}$O (1--0) emission (Fuller et al. 1991) towards the dark cloud B5 region. They found that the two molecular lines showed a similar spatial distribution with both peaks at the same position. One thing H$_2$CO and C$^{18}$O have in common is that they both have been probed to trace the dense gas. The 1$_{10}$--1$_{11}$ transition of H$_2$CO is a unique probe to trace high-density gases at low temperature regions. Liszt \& Lucas (1995) and Liszt et al. (2006) have found that the N(H$_2$CO) are quite well correlated with N(HCO${^+}$), N(C$_2$H), N(HCN), N(CS) and N(NH$_3$) in the diffuse clouds. Toward the cold, dense pre-protostellar cores, the H$_2$CO shows depletion (Young et al. 2004). Therefore, the H$_2$CO will show an interesting difference towards the warm, high-density regions.

In this paper, we report present comparison of H$_2$CO, C$^{18}$O and continuum on a scale of 2 $\sim$ 10 pc towards four galactic HII regions of MON R2, S156, DR17/L906 and M17/M18. We are interested in two issues: (1) to seek the relation between the H$_2$CO line and the background continuum; (2) to make a comparative study on the H$_2$CO and C$^{18}$O lines.

\section{Observations}
\label{sect:Obs}
The $^{12}$CO, $^{13}$CO and C$^{18}$O observations in the four regions of MON R2 (60$'$ $\times$ 90$'$), S156 (50$'$ $\times$ 70$'$), DR17/L906 (40$'$ $\times$ 60$'$) and M17/M18 (70$'$ $\times$ 80$'$) have been carried out at the 13.7 m radio telescope of Purple Mountain Observatory in Delingha from 15 to 26 May 2011. And the $^{12}$CO and $^{13}$CO data have been reported in Tang et al. (2013). The HPBW was 60$''$ at 110 GHz and the beam efficiency was 48\%. The Fast Fourier Transform Spectrometer was used, and the three CO lines were observed simultaneously. The $^{12}$CO velocity resolution was 0.16 km s$^{-1}$ while the velocity resolution of $^{13}$CO and C$^{18}$O was 0.17 km s$^{-1}$. The observation was performed in the On-The-Fly mode. The average integration time of every point was one minute. H$_2$CO and continuum data were selected from Tang et al. (2013). Four HII regions were observed at 1$_{10}$--1$_{11}$ transition of H$_2$CO and 4.8 GHz continuum by Tang et al. (2013) using the Nanshan 25 m radio telescope of Xinjiang Astronomical Observatory.

\section{Results}
\subsection{Data reduction and exhibition}
\label{sect:data}

C$^{18}$O data were reduced using CLASS and GREG which are parts of the GILDAS software developed by IRAM\footnote{%
  \tiny
GILDAS package was developed by IRAM (Institute de Radioastronomie Millim\'{e}trique). http://www.iram.fr/IRAMFR/GILDAS.}. CLASS was used to remove baselines, average weighted spectra, and calibrate the data. The line-center velocities (V$_{Lsr}$) and line width ($\Delta$V) were determined by fitting a manifold of Gaussian profiles. For the comparison of C$^{18}$O and H$_2$CO data, we have smoothed the C$^{18}$O observations to 10$'$, and resampled them on H$_2$CO observing grid. GREG was used to map C$^{18}$O data.

Towards four regions, the C$^{18}$O line spectra are shown in Fig.B.1 and we map the integrated intensity of C$^{18}$O emission in Fig.2. The parameters C$^{18}$O are listed in Table B.1. The optical depth and column density of C$^{18}$O were estimated following Sato et al. (1994) calculations, on the assumption that the cloud is in LTE (Local Thermodynamic Equilibrium) and the excitation temperature of C$^{18}$O is same as that of $^{12}$CO. The H$_2$CO peak optical depth and column density estimations were done following Pipenbrink \& Wendker (1988), using a simply standard radiative transfer result.

\subsection{Description of sources}
MON R2. -- The Mon R2 reflection association (van den Bergh 1966; Racine 1968) is a nearby complex star forming region and is about 830 pc from the Sun (Herbst \& Racine 1976). The molecular content of this region has been the subject of several observational studies for the last several decades. Continuum observations at 6 cm have been reported by Wood \& Churchwell (1989). It shows that the ultracompact HII region is highly asymmetric and reaches its maximum toward its exciting star Mon R2 IRS1. Downes et al. (1975) have found the existence of OH maser emission, H$_2$CO and OH absorption in this region. Using the $^{13}$CO and C$^{18}$O maps, the mass of molecular gas in the Mon R2 core has been determined by Ridge et al. (2003). The C$^{18}$O result shows a gas mass of 1826 M$_{\odot}$, while the more abundant $^{13}$CO gives a mass of 2550 M$_{\odot}$.

Maps of the integrated intensities of the C$^{18}$O line velocities range from 0 to 20 km s$^{-1}$ in Fig.2 (a). The spectrum of intensity peaks of C$^{18}$O shows a velocity components at 10.4 km s$^{-1}$, which agrees with that of the H$_2$CO lines velocity components at 10.5 km s$^{-1}$. Owing to its weakness, the H110$\alpha$ recombination line was not detected by Tang et al. (2013). The formation of massive stars in this region is occurring on the back side of the cloud (Gilmore 1980). It also could be the foreground gas and dust blotting the H110$\alpha$ recombination line. A good agreement between the H$_2$CO and the continuum distributions (see Fig.A.5 (a), Tang et al. 2013) suggests that the continuum emission comes from the HII region. And the H$_2$CO absorbs the background continuum.

S156. -- S156, also known as IC1470, is a compact HII region associated with an extensive molecular cloud lying in the direction of Perseus arm. The 6-cm H$_2$CO absorption and OH emission have been observed by Hoglund \& Gordon (1973) in this region. They found a general correlation among HI absorption, H$_2$CO absorption and OH emission. High resolution continuum observations at 6 cm toward the S156 HII region have been reported by Israel (1977). Its distance is $\sim$ 3.5 -- 4.3 kpc from the Sun (Hoglund \& Gordon, 1973).

The integrated intensities of the C$^{18}$O maps line velocities range from -55 to -45 km s$^{-1}$ in Fig.2 (b). The spectrum of intensity peaks of C$^{18}$O shows a velocity components at -52.1 km s$^{-1}$, which agrees with that of the H$_2$CO at -50.2 km s$^{-1}$. The H110$\alpha$ recombination line was not detected (Tang et al. 2013). The strong absorption extends well outside the continuum source, and is absent in front of weaker continuum emission to the north and south of the main continuum source (see Fig.A.5 (b), Tang et al. 2013). The wider H$_2$CO line width in the central part of the cloud shows that the cloud could be affected by the HII region (Tang et al. 2013). Therefore, the H$_2$CO absorption could absorb the background continuum.

DR17/L906. -- DR17 is an extended HII region with an arc of mid-IR emission associated with the molecular clouds in the Cyg X-North (Schneider et al. 2006). A high resolution continuum at 6-cm survey of the Cygnus X region has been presented by Wendker (1984). The H110$\alpha$ and H$_2$CO at 6-cm have been reported by Pipenbrink \& Wendker (1988). The distance is about 0.8 kpc from the Sun. The source at the southern tip of DR17 is L906, which is a small dark cloud (Davis et al. 2007).

Maps of the integrated intensities of the C$^{18}$O line velocities range from -10 to 20 km s$^{-1}$ in Fig.2 (c). The C$^{18}$O line was not detected towards the DR17 HII region. The C$^{18}$O velocity components at 14.4 km s$^{-1}$ and the H$_2$CO velocity component at 15.4 km s$^{-1}$ are associated with the L906 region. The strong H110$\alpha$ emission was detected in the DR17 region by Tang et al. (2013). The H110$\alpha$ is coincident with the south continuum source, and the H$_2$CO distributions have good agreement with the continuum distribution (see Fig.A.5 (c), Tang et al. 2013). Therefore, the H$_2$CO absorbs the continuum which comes from the DR17 HII region. The agreement between continuum and H$_2$CO distributions in L906 region is not better than that of DR17 region, and the H110$\alpha$ as well as continuum emission are weak. The H$_2$CO may absorb the background continuum in the L906 region.

Towards the DR17 region, the cloud shows a large area of H110$\alpha$ emission (see Fig.A.5 (c), Tang et al. 2013). These imply that there could be a strong source of UV radiation in the DR17 region. 2MASS images have recognized two OB clusters [LK2002]Cl-12 ([DB2001]Cl-15) which could be the exciting sources for the HII region and the mid-IR loop emission (Schneider et al. 2006). The molecular cloud could be eroded by the strong background radiation field, stellar wind or ionized gas in the DR17 HII region. So it shows the comparison between the lower abundance of H$_2$ in the DR17 HII region and the L906 region. Generally, the H$_2$CO absorption line intensity depends on the background continuum temperature and the gas density, so it would be strong in the strong background continuum although the gas density is lower. Only one H$_2$CO peak is associated with continuum emission in the DR17 region. Comparing to H$_2$CO peaks in DR17 region, there are weak continuum and H110$\alpha$ emission associated with H$_2$CO peaks in the L906 region, however, the intensity of H$_2$CO is similar in these two regions. DR17 region has stronger continuum temperature and lower n(H$_2$) density, while weaker continuum emission and higher n(H$_2$) density are in L906 region. Therefore, the reason of similar H$_2$CO intensities in DR17 and L906 regions may be that the different continuum temperatures and different gas densities influence the H$_2$CO absorption line intensity together.

M17/M18. -- M17 features a visible HII region, and it is one of the youngest as well as the most massive nearby star formation regions in the Galaxy (Povich et al. 2009). Observations of 6-cm H$_2$CO toward M17 region have been done by Lada \& Chaisson (1975) and Bieging et al. (1982). The continuum radiation of M17 was mapped at a wavelength of 6 cm by Mezger \& Henderson (1967). The distance is about 1.5 kpc from the Sun (Downes et al.1980). M18 is an open and young cluster in Sagittarius (Lindoff 1971, McSwain \& Gies 2005). It is about 1.25 kpc from the Sun (Lindoff 1971). There are hundreds of dark clouds in M17/M18 regions.

The integrated intensities of the C$^{18}$O maps line velocities range from 10 to 50 km s$^{-1}$ in Fig.2 (d). The spectrum of intensity peaks of C$^{18}$O shows a velocity components at 20.0 km s$^{-1}$, which agrees with that of the H$_2$CO lines at 22.2 km s$^{-1}$. Strong H110$\alpha$ and continuum emission were detected in the M17 region (Tang et al. 2013), and there was a good agreement between them. There is a deviation between H$_2$CO intensity and H110$\alpha$, continuum peaks in the direction of M17; it is about 10$'$ ($\sim$ 4 pc) (see Fig.A.5 (d), Tang et al. 2013). The H$_2$CO absorbs the continuum which comes from the M17 HII region in despite of an offset. At the same time, the peaks of the $^{12}$CO (1--0), $^{13}$CO (1--0) and C$^{18}$O (1--0) intensities are associated with H$_2$CO peaks (see Fig.2 (d) and Figs.A.1 (d) and A.3 (d) in Tang et al. 2013). The M17 HII region erupts from the side of the giant molecular cloud M17 SW (Lada 1976). Tsivilev \& Krasnov (1999) and Pellegrini et al. (2007) showed that the entire HII region in the M17 region is expanding, and the HII region is eroding the M17 molecular cloud. Therefore, it may be because the strong radiation field coming from the M17 HII region ionized the molecular cloud and then affected the giant molecular cloud distribution. This could be a reason for the offset between the H$_2$CO intensity and continuum peaks.

\section{Discussion}
\subsection{Comparison between H$_2$CO and background continuum}
The relation between H$_2$CO line temperatures and continuum temperatures have been noted by Zhang et al. (2012) in four HII regions. They concluded that the H$_2$CO integration intensity is positively correlated with the background continuum temperature when the continuum temperature is below 6 K. Those regions and the four regions Tang et al. (2013) observed have a low range of continuum temperature (T$_C$ $<$ 10 K), thus it is difficult to know the varied process of the H$_2$CO intensity in the high brightness continuum region. So we selected some sources (including giant molecular clouds (GMCs) and HII sources) from the previous observations of H$_2$CO and continuum, and plotted these sources together in Fig.1.  A horizontal slice in figure shows a U shape of the H$_2$CO intensity. The absorption line intensity is proportional to background continuum intensity as the continuum intensity is below about 10 K, which agrees with the conclusion of Zhang et al. (2012). Above this value, the intensity of H$_2$CO line shows slight dispersion, and may weakly correlate with the background continuum intensity. It suggests that the H$_2$CO intensity is strongly influenced by the strong background continuum emission. Actually, H$_2$CO absorption line intensities depend on the background continuum level, but they also depend on H$_2$CO column density and gas density (e.g., Garrison et al. 1975; Wootten et al. 1980).

\begin{figure}[t]
\vspace*{2mm}
\begin{center}
\includegraphics[height=5.5cm]{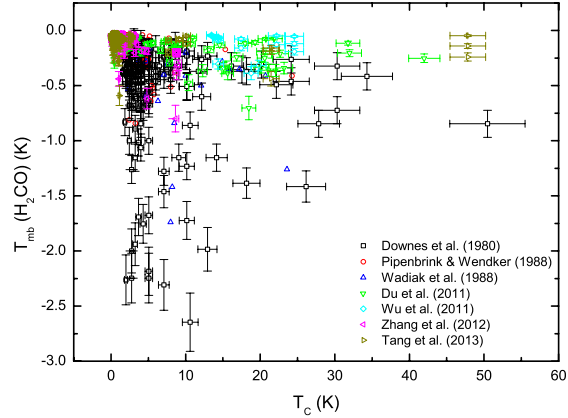}
\end{center}
\caption{The relation between the brightness temperature of H$_2$CO and background continuum.}
\end{figure}

We compared the Non--LTE of interstellar line spectra model (van der Tak et al. 2007) results with the observed results. The Non--LTE model shows similar distribution of the relation between the H$_2$CO line intensity and the background continuum intensity (see Fig.A.1). The absorption line intensity is proportional to the background continuum intensity, which guarantees a rough proportionality between 6 cm continuum and H$_2$CO intensity until the 6 cm continuum reaches some critical values corresponding to the background continuum brightness of about 3 -- 8 K. Above the critical value, the intensity of H$_2$CO line decreases as the background continuum temperature increases. Under this condition where the H$_2$CO cloud has a strong background continuum emission field, the H$_2$CO cloud can also be heated, so part of H$_2$CO cloud could be thermalized. The excitation temperature of H$_2$CO absorption line would increase above 2.7 K. This suggests that the strong background continuum emission affects the excitation of the H$_2$CO absorption line. It could be a reason why the strong H$_2$CO absorption were quite away from the bright continuum peaks in some clouds.

\begin{figure}[t]
\vspace*{2mm}
\begin{center}
\includegraphics[width=10.2cm,angle=0]{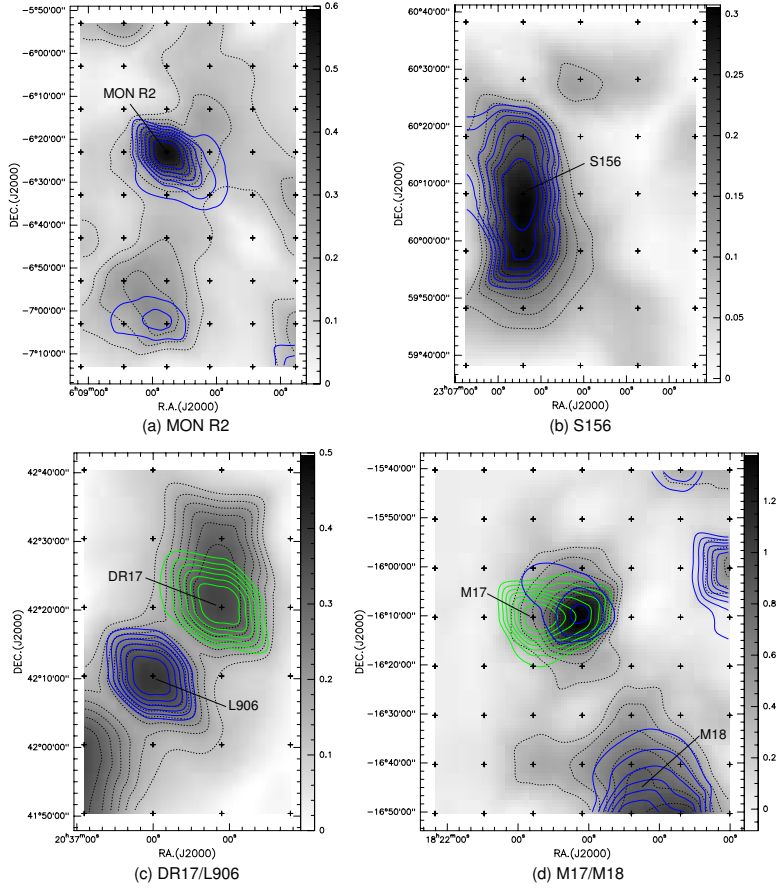}
\end{center}
\caption{Contour maps of integrated area toward (a) MON R2, (b) S156, (c) DR17/L906 and (d) M17/M18. The black contour, green contour and blue contour respectively indicates the integrated intensity of the H$_2$CO absorption line, H110$\alpha$ recombination line and C$^{18}$O emission line. H$_2$CO and H110$\alpha$ data were selected from Tang et al. (2013). (a) For the MON R2 region: C$^{18}$O contour levels are 0.20 to 0.66 in steps of 0.06 K km s$^{-1}$. (b) For the S156 region: C$^{18}$O contour levels are 0.20 to 0.46 in steps of 0.05 K km s$^{-1}$. (c) For the DR17/L906 region: C$^{18}$O contour levels are 0.22 to 0.49 in steps of 0.05 K km s$^{-1}$. (d) For the M17/M18 region: C$^{18}$O contour levels are 0.80 to 2.12 in steps of 0.27 K km s$^{-1}$. The gray bars are given in units of K km s$^{-1}$ for the negative integrated intensity of H$_2$CO.}
\end{figure}

The statistic of observed results shows that approximately 75\% of observed positions with the line-to-continuum ratio $|$T$_L$/T$_C$$|$ are less than 0.2 and the average value $<$$|$T$_L$/T$_C$$|$$>$ is about 0.066 which is higher than the value 0.052 estimated by Whiteoak \& Gardner (1974) for 280 galactic radio sources or source components. Towards the MON R2, S156 and L906 regions the $|$T$_L$/T$_C$$|$ ratios are higher than those towards the DR17 and M17 HII regions. Tang et al. (2013) detected strong H110$\alpha$ and continuum emission in DR17 and M17 regions, but did not detect H110$\alpha$ emissions towards MON R2, S156 and L906 regions which have weak continuum emission. It is suggested that the presence of strong H110$\alpha$ and 6 cm continuum emissions is the reason why high $|$T$_L$/T$_C$$|$ ratios do not occur in these regions.

\subsection{Comparison between H$_2$CO absorption and C$^{18}$O emission}
We compared the velocities of H$_2$CO and C$^{18}$O, and there is a good agreement between two molecules towards four regions (see Table B.1). The distributions of H$_2$CO and C$^{18}$O towards four regions are displayed as a line integral intensity maps in Fig.2, which shows that H$_2$CO and C$^{18}$O have quite a similar geometry and peak positions. The H$_2$CO absorption maximum is located at the same position with C$^{18}$O peak except DR17 region. The map shows that the extent of the H$_2$CO absorption is broader and smaller for the C$^{18}$O emission in four regions. This indicates that the two molecules are probing different parts of the molecular clouds. The C$^{18}$O emission is more likely tracing denser gas than the H$_2$CO absorption is. 176 H$_2$CO absorption points were detected in four regions, but we only detected about 52 points with the C$^{18}$O emission. The C$^{18}$O is somewhat more difficult to be detected than the H$_2$CO is. Those maybe because the density of the H$_2$CO absorption line is lower than that required to excite the C$^{18}$O. The comparison between $^{12}$CO and $^{13}$CO mapping results towards four regions (Tang et al. 2013), shows that $^{12}$CO, $^{13}$CO, H$_2$CO and C$^{18}$O distribution ranges are gradually decreasing. The correlation between the H$_2$CO and the $^{13}$CO distributions is better than that between the H$_2$CO and $^{12}$CO, C$^{18}$O distributions.

\begin{figure}[t]
\vspace*{2mm}
\begin{center}
\includegraphics[height=5.5cm]{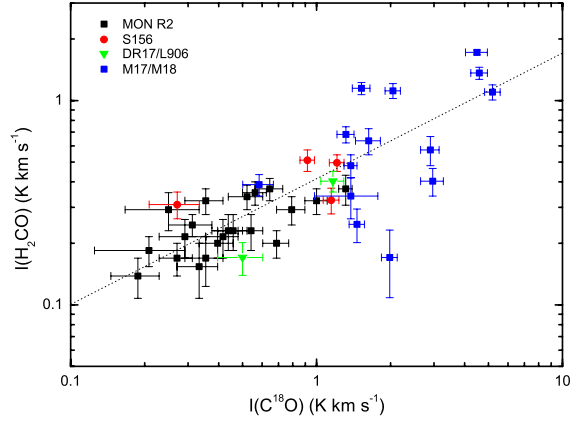}
\end{center}
\caption{Correlation between the H$_2$CO and C$^{18}$O lines fluxes. The dashed line is the linear fit for H$_2$CO and C$^{18}$O flux data.}
\end{figure}

\begin{figure}[h]
\vspace*{2mm}
\begin{center}
\includegraphics[height=5.5cm]{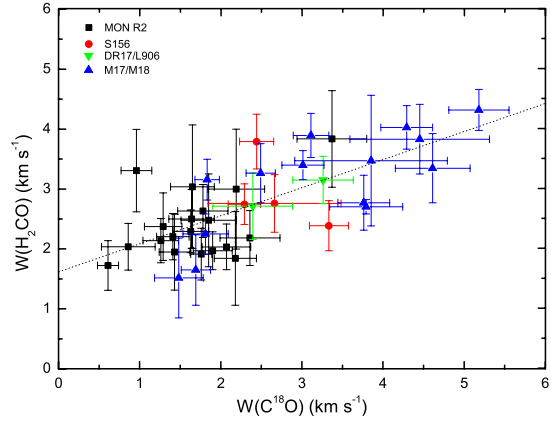}
\end{center}
\caption{Figure shows the line widths of H$_2$CO and C$^{18}$O. The dashed line is the linear fit for H$_2$CO and C$^{18}$O line widths, and the correlation coefficient is 0.7.}
\end{figure}

The relation of fluxes between H$_2$CO and C$^{18}$O was represented. We selected the fluxes data shown as squares in Fig.3, and used them to make linear fits. The best linear fit to a straight line is,
{\setlength\arraycolsep{2pt}
\begin{eqnarray}
 \log I(H_2CO)=(0.60\pm0.06)\log I(C^{18}O)-(0.38\pm0.02) (K km s^{-1}).
\end{eqnarray}
It shows that I(H$_2$CO) is linearly well correlated with I(C$^{18}$O), and the correlation coefficient is 0.8. The H$_2$CO absorption line detections depend on the background continuum brightness and the gas density. For the C$^{18}$O lines, their intensities are dominated by the kinetic temperature of the gas. The fitted relation between I(H$_2$CO) and I(C$^{18}$O) indicates that H$_2$CO absorption line can trace the dense and warm region of the molecular cloud.

The line widths show the thermal motion and turbulence of internal kinematical properties of molecular clouds. H$_2$CO and C$^{18}$O molecules have the same molecular weight, so the thermal line width is about 0.2 km s$^{-1}$ for two molecules, which is concluded at 30 K. All H$_2$CO and C$^{18}$O line widths observed at large scale exceed the thermal line width. In addition, the H$_2$CO line has hyperfine structure components. The contribution of hyperfine structure and thermal broadening to the measured H$_2$CO line widths is likely to be small to moderate (Tang et al. 2013). The correlation between H$_2$CO and $^{12}$CO, $^{13}$CO line widths are not obvious in these four regions (Tang et al. 2013). Towards the line widths between H$_2$CO and C$^{18}$O, it shows a good correlation (see Fig.4). And there is a similar line widths range of 1 -- 5 km s$^{-1}$ for two tracers. These indicate that the line broaden mechanism of H$_2$CO and C$^{18}$O lines could be similar in the dense regions.

\begin{figure}[t]
\vspace*{2mm}
\begin{center}
\includegraphics[height=5.5cm]{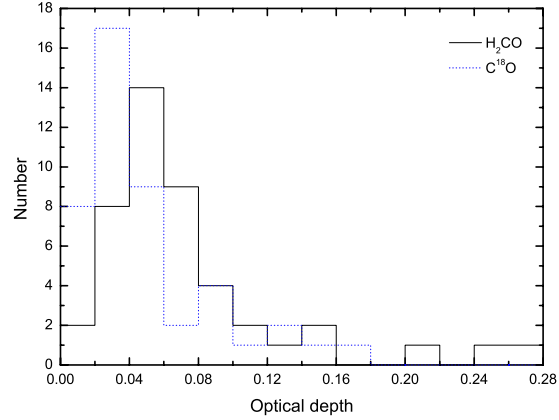}
\end{center}
\caption{Histogram shows the histogram of the peak optical depths of H$_2$CO and C$^{18}$O distribution.}
\end{figure}

\begin{figure}[h]
\vspace*{2mm}
\begin{center}
\includegraphics[height=5.5cm]{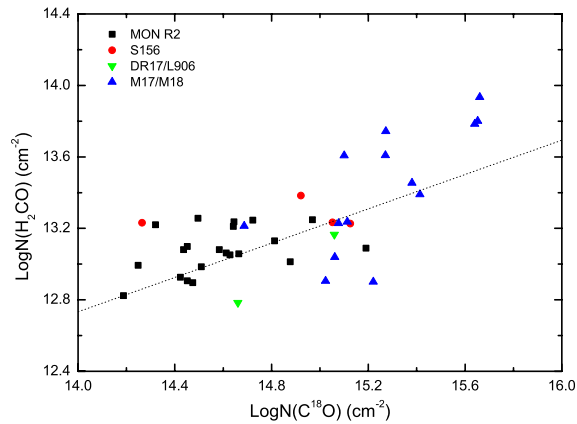}
\end{center}
\caption{Correlation of H$_2$CO and C$^{18}$O column density. The dashed line is the linear fit for H$_2$CO and C$^{18}$O column density, and the correlation coefficient is 0.7.}
\end{figure}

The derived peak optical depths of H$_2$CO and C$^{18}$O show that most of the H$_2$CO optical depths are slightly higher than those of C$^{18}$O (see Fig.5). The average optical depth of H$_2$CO is $<$$\tau$(H$_2$CO)$>$ $\sim$ 0.05 where the C$^{18}$O also have been detected. For C$^{18}$O, the average optical depth is $<$$\tau$(C$^{18}$O)$>$ $\sim$ 0.03. The similar average optical depths of C$^{18}$O and H$_2$CO indicate that the two tracers can both probe high density regions. The relation between H$_2$CO and C$^{18}$O peak column densities shows that the H$_2$CO has a rapid increase in abundance and is not as rapid as that of C$^{18}$O (see Fig.6). For the four regions, N(H$_2$CO) is well correlated with N(C$^{18}$O) and the average column density ratio is $<$N(H$_2$CO)/N(C$^{18}$O)$>$ $\sim$ 0.03. The column density of H$_2$CO may be underestimated because the cloud size is smaller than our beam size. The true cloud size is overestimated (see e.g. Gardner \& Whiteoak, 1972; Slysh,1975), therefore the column density N(H$_2$CO) may be higher than we estimated. At the same time, the column density of C$^{18}$O could be underestimated too.

\section{Summary}
We have carried out large scale C$^{18}$O (1--0) mapping observation towards four galactic HII regions using the 13.7 m radio telescope of Purple Mountain Observatory. Four regions of MON R2, S156, DR17/L906 and M17/M18 have been compared in the transitions of H$_2$CO (1$_{10}$--1$_{11}$), C$^{18}$O (1--0) and 6 cm continuum. The comparison of study would be useful for the scientific community to investigate molecular material associated with HII regions. The main conclusions of this study are as shown in the following:

The H$_2$CO integrated intensity distributions are similar to those of the 6 cm continuum towards MON R2 region. This suggests that the H$_2$CO intensity is strongly influenced by the background continuum emission. In S156 region, the H$_2$CO cloud might be affected by the HII region. The reason for similar H$_2$CO intensities towards DR17 and L906 regions could be that the continuum temperatures and the gas densities influence the H$_2$CO absorption line intensities together. Towards M17, there is a 10$'$ offset between the H$_2$CO absorption and the continuum emission intensity peaks.

The analysis of the observed data and the Non--LTE model results show that the H$_2$CO absorption line intensity is not proportional to background continuum intensity when the continuum intensity is above about 10 K. And the Non--LTE model shows that the brightness temperature of the H$_2$CO absorption line is the strongest in the background continuum temperature range of about 3 -- 8 K. The excitation of the H$_2$CO absorption line is affected by strong background continuum emissions. Approximately 75\% of observed positions with the line-to-continuum ratio $|$T$_L$/T$_C$$|$ are less than 0.2 and the average value $<$$|$T$_L$/T$_C$$|$$>$ $\sim$ 0.066.

From the comparison of H$_2$CO and C$^{18}$O maps, the extent of H$_2$CO absorption is broader than that of C$^{18}$O emission in the four regions. Except for the DR17 region, the H$_2$CO absorption maximum is located at the same position as the C$^{18}$O peak. The intensities and widths of H$_2$CO absorptions and C$^{18}$O emission lines correlate fairly well. This indicates that the H$_2$CO absorption line can trace dense and warm regions of the molecular cloud. For the four regions, N(H$_2$CO) is well correlated with N(C$^{18}$O) and the average column density ratio is $<$N(H$_2$CO)/N(C$^{18}$O)$>$ $\sim$ 0.03.

\begin{acknowledgements}
We thank Zhi. Bo. Jiang, Zhi. Wei. Chen and Jun. Yu. Li of Purple Mountain Observatory for providing the CO data of M17/M18 region. And we thank Shu. Fei. Yu for help with our English expressions. This work was funded by The National Natural Science foundation of China under grant 10778703 and partly supported by China Ministry of Science and Technology under State Key Development Program for Basic Research (2012CB821800) and the National Natural Science foundation of China under grant 11373062, 11303081 and 10873025.
\end{acknowledgements}

\appendix                  

\section{Non-LTE model for H$_2$CO}
The model is based on statistical equilibrium calculations which involve collisional and radiative processes and include radiation from background sources, and the optical depth effects are treated with an escape probability method. The Non--LTE model was generated by assuming that the hydrogen density n(H$_2$) = 10$^4$ cm$^{-3}$ (Garrison et al. 1975, Green et al. 1978, Rodr\'{\i}guez et al. 2007, Troscompt et al. 2009), H$_2$CO column density N(H$_2$CO) = 1.0$\times$10$^{13}$, 5.0$\times$10$^{13}$, and 1.0$\times$10$^{14}$ cm$^{-2}$ (Federman et al. 1990, Troscompt et al. 2009, Tang et al. 2013), H$_2$CO line width FWHM $\Delta$V(H$_2$CO) = 2.5 km s$^{-1}$ (Pipenbrink \& Wendker 1988, Tang et al. 2013) and the kinetic temperatures T$_k$ range is 10 -- 40 K.

\begin{figure}[h]
\vspace*{2mm}
\begin{center}
\includegraphics[height=15cm, angle=-90]{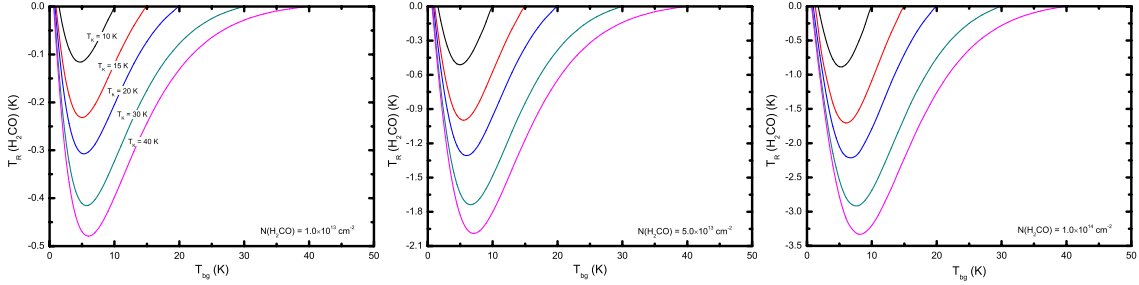}
\end{center}
\caption{Brightness temperature of H$_2$CO vs. background continuum for different kinetic temperatures T$_k$. The plot was generated assuming hydrogen density n(H$_2$) = 10$^4$ cm$^{-3}$; H$_2$CO column density N(H$_2$CO) = 1.0$\times$10$^{13}$, 5.0$\times$10$^{13}$, and 1.0$\times$10$^{14}$ cm$^{-2}$; line width FWHM $\Delta$V(H$_2$CO) = 2.5 km s$^{-1}$ and the kinetic temperatures T$_k$ = 10, 15, 20, 30, and 40 K, using Non-LTE of interstellar line spectra model code (van der Tak et. al 2007).}
\end{figure}

\section{The line parameters and spectra of C$^{18}$O}

   \begin{longtable}{lcccccccc}
    \caption{The parameters of C$^{18}$O(1--0) and H$_2$CO (1$_{10}$--1$_{11}$)} \\
     \hline\hline
  Sources & Offset & I(C$^{18}$O) & Velocity & Width & T$_A$$^{\ast}$ & I(H$_2$CO) & Velocity & Included in \\
          &(arcmin)& (K km s$^{-1}$) & (km s$^{-1}$) & (km s$^{-1}$) & (K) & (K km s$^{-1}$) & (km s$^{-1}$) & fits \\
 \hline
  \endfirsthead
   \caption{continued.}\\
    \hline\hline
Sources & Offset & I(C$^{18}$O) & Velocity & Width & T$_A$$^{\ast}$ & I(H$_2$CO) & Velocity & Included in \\
          &(arcmin)& (K km s$^{-1}$) & (km s$^{-1}$) & (km s$^{-1}$) & (K) & (K km s$^{-1}$) & (km s$^{-1}$) & fits \\
 \hline
  \endhead
   \hline
    \endfoot

MON R2    &   0,    0 &                 &              &                &       &  -0.30  (0.04) &  7.6   (0.2) &    \\
          &           &    0.63 (0.04)  &  10.4  (0.1)  & 1.76  (0.13)  &  0.34 &  -0.24  (0.04) &  10.5  (0.1) &  Y \\
          & -10,    0 &    0.48 (0.05)  &  10.7  (0.2)  & 3.37  (0.43)  &  0.13 &  -0.21  (0.03) &  10.7  (0.3) &  Y \\
          & -10,  -10 &    0.33 (0.04)  &  10.5  (0.2)  & 2.36  (0.37)  &  0.13 &  -0.13  (0.02) &  10.6  (0.2) &  Y \\
          &   0,  -10 &    0.22 (0.03)  &  10.4  (0.1)  & 1.64  (0.28)  &  0.12 &  -0.15  (0.03) &  9.7   (0.3) &  Y \\
          &  10,  -10 &    $<$0.16      &               &               &       &  -0.08  (0.02) &  9.3   (0.1) &    \\
          &  10,   0  &    0.21 (0.03)  &  9.7   (0.1)  & 1.43  (0.19)  &  0.14 &  -0.15  (0.03) &  9.1   (0.2) &  Y \\
          &  10,   10 &    $<$0.06      &               &               &       &  -0.05  (0.02) &  9.9   (0.2) &    \\
          &   0,   10 &    $<$0.08      &               &               &       &  -0.09  (0.02) &  10.5  (0.1) &    \\
          & -10,   10 &    0.17 (0.03)  &  10.7  (0.1)  & 0.96  (0.19)  &  0.16 &  -0.21  (0.03) &  10.7  (0.3) &  Y \\
          & -20,   10 &    0.17 (0.03)  &  11.8  (0.2)  & 1.65  (0.27)  &  0.10 &  -0.11  (0.03) &  11.9  (0.5) &  Y \\
          & -20,    0 &    0.14 (0.03)  &  12.1  (0.3)  & 2.19  (0.35)  &  0.06 &  -0.14  (0.03) &  11.6  (0.4) &  Y \\
          & -20,  -10 &    0.13 (0.02)  &  11.1  (0.1)  & 1.29  (0.23)  &  0.10 &  -0.11  (0.02) &  10.9  (0.3) &  Y \\
          & -20,  -20 &    $<$0.05      &               &               &       &   ...          &              &    \\
          & -10,  -20 &    $<$0.09      &               &               &       &  -0.09  (0.03) &  10.1  (0.6) &    \\
          &   0,  -20 &    0.15 (0.03)  &  10.2  (0.1)  & 1.33  (0.32)  &  0.11 &  -0.16  (0.02) &  9.1   (0.1) &  Y \\
          &  10,  -20 &    0.09 (0.02)  &  10.4  (0.1)  & 0.61  (0.13)  &  0.14 &  -0.09  (0.02) &  10.0  (0.3) &  Y \\
          &  20,  -20 &    $<$0.03      &               &               &       &  -0.12  (0.03) &  9.8   (0.3) &    \\
          &  20,  -10 &    $<$0.07      &               &               &       &  -0.09  (0.03) &  9.2   (0.4) &    \\
          &  20,    0 &    $<$0.04      &               &               &       &  -0.08  (0.02) &  9.0   (0.3) &    \\
          &  20,   10 &    $<$0.09      &               &               &       &   ...          &              &    \\
          &  20,   20 &    $<$0.05      &               &               &       &   ...          &              &    \\
          &  10,   20 &    $<$0.10      &               &               &       &   ...          &              &    \\
          &   0,   20 &    0.10 (0.04)  &  10.7  (0.2)  & 0.86  (0.33)  &  0.11 &  -0.12  (0.02) &  9.9   (0.2) &  Y \\
          & -10,   20 &    0.20 (0.03)  &  11.7  (0.2)  & 2.18  (0.26)  &  0.09 &  -0.14  (0.03) &  11.1  (0.2) &  Y \\
          & -20,   20 &    0.16 (0.03)  &  12.4  (0.2)  & 1.85  (0.34)  &  0.08 &  -0.10  (0.03) &  12.0  (0.4) &  Y \\
          & -30,   20 &    $<$0.08      &               &               &       &   ...          &              &    \\
          & -30,   10 &    $<$0.09      &               &               &       &  -0.07  (0.03) &  13.6  (0.4) &    \\
          & -30,    0 &    $<$0.04      &               &               &       &  -0.10  (0.02) &  11.9  (0.2) &    \\
          & -30,  -10 &    $<$0.10      &               &               &       &  -0.21  (0.03) &  11.8  (0.3) &    \\
          & -30,  -20 &    $<$0.09      &               &               &       &  -0.10  (0.03) &  8.5   (0.4) &    \\
          & -30,  -30 &    ...          &               &               &       &  -0.14  (0.05) &  9.2   (0.5) &    \\
          & -20,  -30 &    ...          &               &               &       &  -0.07  (0.03) &  9.8   (0.4) &    \\
          & -10,  -30 &    ...          &               &               &       &  -0.09  (0.04) &  9.5   (0.9) &    \\
          &   0,  -30 &    0.25 (0.04)  &  10.5  (0.1)  & 1.78  (0.31)  &  0.13 &  -0.22  (0.03) &  8.9   (0.2) &  Y \\
          &  10,  -30 &    0.27 (0.03)  &  10.1  (0.1)  & 2.07  (0.30)  &  0.12 &  -0.23  (0.03) &  9.6   (0.1) &  Y \\
          &  20,  -30 &    $<$0.01      &               &               &       &  -0.11  (0.03) &  8.9   (0.3) &    \\
          &  20,   30 &    $<$0.07      &               &               &       &  -0.14  (0.05) &  9.6   (0.6) &    \\
          &  10,   30 &    $<$0.12      &               &               &       &  -0.19  (0.04) &  10.0  (0.4) &    \\
          &   0,   30 &    ...          &               &               &       &  -0.08  (0.02) &  10.1  (0.3) &    \\
          & -10,   30 &    0.19 (0.03)  &  12.4  (0.1)  & 1.26  (0.22)  &  0.14 &  -0.13  (0.02) &  11.9  (0.1) &  Y \\
          & -20,   30 &    0.10 (0.03)  &  12.2  (0.2)  & 1.40  (0.36)  &  0.07 &   ...          &              &    \\
          & -30,   30 &    $<$0.10      &               &               &       &   ...          &              &    \\
          & -30,  -40 &    $<$0.07      &               &               &       &  -0.12  (0.03) &  11.8  (0.3) &    \\
          & -20,  -40 &    $<$0.05      &               &               &       &  -0.05  (0.02) &  12.4  (0.5) &    \\
          & -10,  -40 &    $<$0.06      &               &               &       &  -0.09  (0.02) &  8.4   (0.4) &    \\
          &   0,  -40 &    0.31 (0.04)  &  10.1  (0.1)  & 1.63  (0.22)  &  0.18 &  -0.24  (0.03) &  9.3   (0.1) &  Y \\
          &  10,  -40 &    0.26 (0.03)  &  10.1  (0.1)  & 1.90  (0.24)  &  0.13 &  -0.15  (0.03) &  9.9   (0.2) &  Y \\
          &  20,  -40 &    $<$0.05      &               &               &       &  -0.11  (0.04) &  9.2   (0.2) &    \\
          & -30,  -50 &    0.38 (0.05)  &  12.2  (0.1)  & 1.41  (0.20)  &  0.26 &  -0.19  (0.03) &  12.1  (0.2) &  Y \\
          & -20,  -50 &    $<$0.04      &               &               &       &   ...          &              &    \\
          & -10,  -50 &    $<$0.11      &               &               &       &   ...          &              &    \\
          &   0,  -50 &    ...          &               &               &       &   ...          &              &    \\
          &  10,  -50 &    ...          &               &               &       &  -0.07  (0.03) &  10.3  (0.4) &    \\
          &  20,  -50 &    0.08 (0.02)  &  10.7  (0.1)  & 0.84  (0.23)  &  0.09 &   ...          &              &    \\
S156      &   0,   0  &    0.58 (0.04)  & -52.1  (0.1)  & 2.44  (0.21)  &  0.23 &  -0.32  (0.03) & -50.2  (0.2) &  Y \\
          & -10,   0  &    $<$0.04      &               &               &       &  -0.09  (0.02) & -49.0  (0.3) &    \\
          & -10,  -10 &    $<$0.05      &               &               &       &  -0.13  (0.03) & -48.7  (0.4) &    \\
          &   0,  -10 &    0.44 (0.03)  & -51.9  (0.1)  & 2.29  (0.20)  &  0.18 &  -0.33  (0.04) & -50.2  (0.2) &  Y \\
          &  10,  -10 &    ...          &               &               &       &   ...          &              &    \\
          &  10,   0  &    0.25 (0.06)  & -53.0  (0.2)  & 1.43  (0.45)  &  0.16 &   ...          &              &    \\
          &  10,   10 &    0.23 (0.07)  & -53.0  (0.2)  & 1.20  (0.58)  &  0.18 &  -0.09  (0.03) & -50.1  (0.2) &    \\
          &   0,   10 &    0.55 (0.04)  & -51.1  (0.1)  & 3.33  (0.24)  &  0.15 &  -0.21  (0.03) & -50.2  (0.2) &  Y \\
          & -10,   10 &    $<$0.02      &               &               &       &  -0.08  (0.02) & -49.0  (0.3) &    \\
          & -20,   10 &    $<$0.02      &               &               &       &  -0.05  (0.02) & -45.0  (0.4) &    \\
          & -20,   0  &    ...          &               &               &       &   ...          &              &    \\
          & -20,  -10 &    $<$0.08      &               &               &       &   ...          &              &    \\
          & -20,  -20 &    $<$0.03      &               &               &       &   ...          &              &    \\
          & -10,  -20 &    ...          &               &               &       &  -0.12  (0.04) & -49.5  (0.6) &    \\
          &   0,  -20 &    0.13 (0.03)  & -52.2  (0.4)  & 2.66  (0.82)  &  0.04 &  -0.20  (0.03) & -49.4  (0.2) &  Y \\
          &  10,  -20 &    $<$0.05      &               &               &       &  -0.09  (0.03) & -50.6  (0.5) &    \\
          &  10,   20 &    $<$0.01      &               &               &       &  -0.08  (0.03) & -50.4  (0.4) &    \\
          &   0,   20 &    $<$0.02      &               &               &       &  -0.07  (0.02) & -49.1  (0.3) &    \\
          & -10,   20 &    $<$0.03      &               &               &       &  -0.13  (0.03) & -50.3  (0.4) &    \\
          & -20,   20 &    $<$0.01      &               &               &       &  -0.10  (0.04) & -51.1  (0.7) &    \\
          & -30,   20 &    ...          &               &               &       &   ...          &              &    \\
          & -30,   10 &    $<$0.03      &               &               &       &  -0.08  (0.02) & -50.2  (0.4) &    \\
          & -30,   0  &    $<$0.02      &               &               &       &   ...          &              &    \\
          & -30,  -10 &    $<$0.01      &               &               &       &  -0.09  (0.04) & -46.8  (0.9) &    \\
          & -30,  -20 &    $<$0.04      &               &               &       &   ...          &              &    \\
          & -30,  -30 &    $<$0.02      &               &               &       &  -0.08  (0.03) & -54.3  (0.4) &    \\
          & -20,  -30 &    $<$0.02      &               &               &       &   ...          &              &    \\
          & -10,  -30 &    ...          &               &               &       &   ...          &              &    \\
          &   0,  -30 &    ...          &               &               &       &   ...          &              &    \\
          &  10,  -30 &    $<$0.01      &               &               &       &   ...          &              &    \\
          &  10,   30 &    $<$0.03      &               &               &       &   ...          &              &    \\
          &   0,   30 &    $<$0.04      &               &               &       &   ...          &              &    \\
          & -10,   30 &    $<$0.02      &               &               &       &   ...          &              &    \\
          & -20,   30 &    $<$0.04      &               &               &       &   ...          &              &    \\
          & -30,   30 &    $<$0.04      &               &               &       &   ...          &              &    \\
DR17/L906 &   0,   0  &    $<$0.05      &               &               &       &  -0.37  (0.05) &  7.0   (0.2) &    \\
          & -10,   0  &    $<$0.04      &               &               &       &  -0.11  (0.02) &  5.8   (0.3) &    \\
          & -10,  -10 &    ...          &               &               &       &  -0.09  (0.03) &  7.7   (0.5) &    \\
          &   0,  -10 &    $<$0.01      &               &               &       &  -0.06  (0.03) &  6.9   (0.3) &    \\
          &           &                 &               &               &       &  -0.09  (0.03) &  9.8   (0.5) &    \\
          &  10,  -10 &                 &               &               &       &  -0.10  (0.02) &  7.1   (0.2) &    \\
          &           &    0.56 (0.06)  &  14.4  (0.2)  & 3.26  (0.34)  &  0.16 &  -0.26  (0.03) &  15.4  (0.2) &  Y \\
          &  10,   0  &    $<$0.11      &               &               &       &  -0.14  (0.02) &  6.5   (0.2) &    \\
          &  10,   10 &    $<$0.05      &               &               &       &  -0.07  (0.01) &  6.6   (0.2) &    \\
          &   0,   10 &    $<$0.04      &               &               &       &  -0.23  (0.03) &  6.2   (0.3) &    \\
          &           &                 &               &               &       &  -0.05  (0.02) &  14.7  (0.4) &    \\
          & -10,   10 &    ...          &               &               &       &  -0.07  (0.02) &  6.8   (0.3) &    \\
          & -10,  -20 &    ...          &               &               &       &  -0.06  (0.03) &  5.9   (0.5) &    \\
          &  0,   -20 &    $<$0.01      &               &               &       &  -0.06  (0.02) &  6.4   (0.3) &    \\
          &           &                 &               &               &       &  -0.05  (0.02) &  10.7  (0.4) &    \\
          &  10,  -20 &    $<$0.06      &               &               &       &  -0.12  (0.02) &  6.5   (0.2) &    \\
          &  20,  -20 &    $<$0.09      &               &               &       &  -0.22  (0.03) &  4.6   (0.2) &    \\
          &           &                 &               &               &       &  -0.07  (0.02) &  14.0  (0.1) &    \\
          &  20,  -10 &                 &               &               &       &  -0.05  (0.02) &  4.6   (0.4) &    \\
          &           &    0.24 (0.05)  &  15.3  (0.3)  & 2.40  (0.49)  &  0.10 &  -0.11  (0.02) &  15.8  (0.2) &  Y \\
          &  20,   0  &    $<$0.03      &               &               &       &   ...          &              &    \\
          &  20,   10 &    $<$0.02      &               &               &       &   ...          &              &    \\
          &  20,   20 &    ...          &               &               &       &   ...          &              &    \\
          &  10,   20 &    $<$0.06      &               &               &       &  -0.11  (0.03) &  5.7   (0.6) &    \\
          &   0,   20 &    $<$0.02      &               &               &       &  -0.08  (0.04) &  9.7   (0.4) &    \\
          & -10,   20 &    $<$0.02      &               &               &       &  -0.09  (0.04) &  7.5   (0.7) &    \\
          & -10,  -30 &    ...          &               &               &       &   ...          &              &    \\
          &  0,   -30 &    $<$0.06      &               &               &       &  -0.05  (0.02) &  5.4   (0.2) &    \\
          &  10,  -30 &    $<$0.04      &               &               &       &  -0.09  (0.02) &  7.4   (0.2) &    \\
          &  20,  -30 &    $<$0.05      &               &               &       &  -0.25  (0.01) &  6.6   (0.3) &    \\
          &           &                 &               &               &       &  -0.05  (0.02) &  14.0  (0.1) &    \\
M17/M18   &  0,   0   &                 &               &               &       &  -0.17  (0.04) &  16.6  (0.1) &    \\
          &           &    $<$0.75      &               &               &       &  -0.36  (0.04) &  22.2  (0.1) &    \\
          &           &                 &               &               &       &  -0.08  (0.04) &  38.1  (0.6) &    \\
          & -10,   0  &                 &               &               &       &  -0.49  (0.06) &  15.0  (1.2) &    \\
          &           &                 &               &               &       &  -0.43  (0.06) &  19.2  (1.2) &    \\
          &           &    0.78 (0.09)  &  20.6  (0.3)  & 6.19  (0.88)  &  0.12 &  -0.41  (0.06) &  22.4  (1.2) &  Y \\
          &           &    0.69 (0.08)  &  31.5  (0.4)  & 5.74  (0.80)  &  0.11 &                &              &    \\
          & -10,  -10 &    $<$0.12      &               &               &       &  -0.32  (0.04) &  18.7  (0.4) &    \\
          &           &                 &               &               &       &  -0.06  (0.03) &  26.2  (0.2) &    \\
          &   0,  -10 &    $<$0.19      &               &               &       &  -0.10  (0.05) &  18.4  (0.5) &    \\
          &           &                 &               &               &       &  -0.10  (0.03) &  21.2  (0.1) &    \\
          &  10,  -10 &    $<$0.37      &               &               &       &   ...          &              &    \\
          &  10,  0   &    $<$0.15      &               &               &       &   ...          &              &    \\
          &  10,   10 &    $<$0.22      &               &               &       &  -0.05  (0.05) &  20.9  (1.2) &    \\
          &           &                 &               &               &       &  -0.06  (0.05) &  22.7  (0.5) &    \\
          &   0,   10 &                 &               &               &       &  -0.24  (0.05) &  17.8  (0.4) &    \\
          &           &    $<$0.56      &               &               &       &  -0.13  (0.04) &  22.3  (0.4) &    \\
          &           &    0.33 (0.07)  &  34.7  (0.5)  & 4.41  (1.04)  &  0.07 &                &              &    \\
          & -10,   10 &    $<$0.51      &               &               &       &  -0.54  (0.06) &  17.5  (0.2) &    \\
          &           &                 &               &               &       &  -0.10  (0.04) &  21.9  (0.4) &    \\
          & -20,   10 &    ...          &               &               &       &  -0.06  (0.02) &  39.3  (0.2) &    \\
          & -20,   0  &    $<$0.24      &               &               &       &  -0.10  (0.02) &  30.2  (0.2) &    \\
          & -20,  -10 &                 &               &               &       &  -0.07  (0.04) &  18.1  (1.1) &    \\
          &           &    0.63 (0.06)  &  28.5  (0.2)  & 3.53  (0.37)  &  0.17 &                &              &    \\
          & -20,  -20 &    0.66 (0.04)  &  18.5  (0.1)  & 2.49  (0.18)  &  0.25 &  -0.31  (0.04) &  17.6  (0.2) &  Y \\
          &           &                 &               &               &       &  -0.07  (0.03) &  28.4  (0.5) &    \\
          &           &                 &               &               &       &  -0.08  (0.04) &  37.0  (0.5) &    \\
          & -10,  -20 &    ...          &               &               &       &  -0.07  (0.02) &  18.1  (0.3) &    \\
          &   0,  -20 &    0.28 (0.04)  &  18.6  (0.1)  & 1.81  (0.28)  &  0.15 &  -0.25  (0.03) &  17.7  (0.1) &  Y \\
          &           &                 &               &               &       &  -0.05  (0.02) &  32.2  (0.2) &    \\
          &  10,  -20 &    $<$0.17      &               &               &       &  -0.06  (0.02) &  19.2  (0.3) &    \\
          &  20,  -20 &    $<$0.04      &               &               &       &  -0.03  (0.03) &  21.8  (0.8) &    \\
          &           &                 &               &               &       &  -0.13  (0.06) &  37.8  (1.3) &    \\
          &  20,  -10 &    $<$0.27      &               &               &       &  -0.06  (0.04) &  15.5  (0.7) &    \\
          &  20,   0  &    $<$0.23      &               &               &       &   ...          &              &    \\
          &  20,   10 &    $<$0.15      &               &               &       &   ...          &              &    \\
          &  20,   20 &    $<$0.15      &               &               &       &   ...          &              &    \\
          &  10,   20 &    $<$0.12      &               &               &       &  -0.09  (0.03) &  21.5  (0.6) &    \\
          &   0,   20 &    $<$0.12      &               &               &       &  -0.16  (0.04) &  19.9  (0.8) &    \\
          & -10,   20 &    ...          &               &               &       &  -0.07  (0.02) &  20.0  (0.1) &    \\
          & -20,   20 &    ...          &               &               &       &  -0.11  (0.03) &  21.3  (0.1) &    \\
          &           &                 &               &               &       &  -0.05  (0.02) &  29.4  (0.4) &    \\
          & -30,   20 &    $<$0.40      &               &               &       &  -0.07  (0.03) &  29.2  (0.6) &    \\
          & -30,   10 &    $<$0.19      &               &               &       &  -0.05  (0.03) &  39.3  (0.3) &    \\
          & -30,   0  &    $<$0.15      &               &               &       &  -0.05  (0.02) &  22.0  (0.3) &    \\
          & -30,  -10 &                 &               &               &       &  -0.06  (0.02) &  23.7  (0.1) &    \\
          &           &                 &               &               &       &  -0.07  (0.02) &  27.3  (0.2) &    \\
          &           &    0.70 (0.05)  &  41.2  (0.1)  & 3.76  (0.32)  &  0.17 &  -0.16  (0.03) &  39.1  (0.2) &  Y \\
          & -30,  -20 &                 &               &               &       &  -0.06  (0.02) &  16.9  (0.2) &    \\
          &           &                 &               &               &       &  -0.15  (0.03) &  20.7  (0.3) &    \\
          &           &    0.70 (0.06)  &  41.8  (0.2)  & 5.37  (0.53)  &  0.12 &                &              &    \\
          & -30,  -30 &    0.98 (0.07)  &  21.1  (0.2)  & 4.29  (0.32)  &  0.22 &  -0.72  (0.06) &  18.2  (0.2) &  Y \\
          &           &                 &               &               &       &  -0.10  (0.05) &  32.6  (0.9) &    \\
          & -20,  -30 &    0.73 (0.06)  &  20.1  (0.1)  & 3.11  (0.22)  &  0.28 &  -0.74  (0.05) &  18.6  (0.1) &  Y \\
          &           &                 &               &               &       &  -0.06  (0.03) &  34.9  (0.6) &    \\
          & -10,  -30 &    0.63 (0.05)  &  20.6  (0.1)  & 1.83  (0.15)  &  0.33 &  -0.44  (0.04) &  18.7  (0.1) &  Y \\
          &           &                 &               &               &       &  -0.11  (0.04) &  35.4  (0.6) &    \\
          &   0,  -30 &    $<$0.34      &               &               &       &  -0.24  (0.06) &  16.8  (0.6) &    \\
          &           &                 &               &               &       &  -0.05  (0.04) &  34.2  (1.2) &    \\
          &  10,  -30 &    $<$0.09      &               &               &       &  -0.08  (0.04) &  19.6  (0.3) &    \\
          &           &                 &               &               &       &  -0.07  (0.04) &  34.4  (0.4) &    \\
          &  20,  -30 &    $<$0.22      &               &               &       &  -0.05  (0.03) &  37.2  (1.0) &    \\
          &  20,   30 &    $<$0.06      &               &               &       &   ...          &              &    \\
          &  10,   30 &    $<$0.07      &               &               &       &  -0.09  (0.05) &  28.9  (0.9) &    \\
          &   0,   30 &    ...          &               &               &       &   ...          &              &    \\
          & -10,   30 &    $<$0.06      &               &               &       &   ...          &              &    \\
          & -20,   30 &    0.55 (0.07)  &  31.4  (0.1)  & 1.86  (0.30)  &  0.28 &   ...          &              &    \\
          & -30,   30 &                 &               &               &       &  -0.08  (0.03) &  23.4  (0.3) &    \\
          &           &    0.95 (0.07)  &  31.1  (0.1)  & 1.69  (0.18)  &  0.53 &  -0.11  (0.04) &  28.1  (0.4) &  Y \\
          &           &                 &               &               &       &  -0.06  (0.04) &  38.3  (0.3) &    \\
          & -40,   30 &    ...          &               &               &       &  -0.14  (0.05) &  26.9  (0.4) &    \\
          & -40,   20 &    $<$0.05      &               &               &       &  -0.16  (0.07) &  27.5  (1.3) &    \\
          & -40,   10 &    2.49 (0.18)  &  28.3  (0.2)  & 5.18  (0.37)  &  0.45 &  -0.71  (0.06) &  25.4  (0.2) &  Y \\
          & -40,   0  &                 &               &               &       &  -0.11  (0.04) &  26.3  (0.2) &    \\
          &           &    1.42 (0.15)  &  39.5  (0.2)  & 3.49  (0.48)  &  0.38 &  -0.26  (0.04) &  36.9  (0.1) &  Y \\
          & -40,  -10 &    1.39 (0.12)  &  40.5  (0.2)  & 4.61  (0.46)  &  0.28 &  -0.37  (0.06) &  37.8  (0.3) &  Y \\
          & -40,  -20 &    $<$0.19      &               &               &       &  -0.10  (0.04) &  18.9  (0.2) &    \\
          &           &                 &               &               &       &  -0.08  (0.05) &  23.9  (0.9) &    \\
          &           &                 &               &               &       &  -0.22  (0.08) &  35.3  (0.8) &    \\
          & -40,  -30 &    $<$0.15      &               &               &       &  -0.11  (0.04) &  34.0  (0.6) &    \\
          & -40,  -40 &    $<$0.12      &               &               &       &  -0.62  (0.07) &  18.3  (0.2) &    \\
          &           &                 &               &               &       &  -0.12  (0.06) &  32.6  (1.0) &    \\
          & -30,  -40 &    2.15 (0.22)  &  20.7  (0.2)  & 3.79  (0.45)  &  0.53 &  -1.11  (0.03) &  18.5  (0.1) &  Y \\
          &           &                 &               &               &       &  -0.08  (0.04) &  36.8  (0.3) &    \\
          & -20,  -40 &    2.20 (0.17)  &  21.2  (0.1)  & 3.01  (0.26)  &  0.69 &  -0.88  (0.06) &  18.7  (0.1) &  Y \\
          &           &                 &               &               &       &  -0.05  (0.03) &  36.5  (0.6) &    \\
          & -10,  -40 &    0.66 (0.19)  &  21.8  (0.6)  & 3.85  (0.94)  &  0.16 &  -0.22  (0.05) &  18.3  (0.4) &  Y \\
          &           &                 &               &               &       &  -0.09  (0.04) &  37.1  (0.7) &    \\
          &   0,  -40 &    ...          &               &               &       &  -0.06  (0.03) &  18.6  (0.6) &    \\
          &  10,  -40 &    ...          &               &               &       &  -0.23  (0.06) &  26.2  (0.1) &    \\
          &  20,  -40 &    $<$0.10      &               &               &       &   ...          &              &    \\
\hline
 \end{longtable}
  \tablecomments{0.86\textwidth}{The coordinates of four regions for the offset (0, 0) position are MON R2 (06$^h$07$^m$46$^s$.60, $-06^{\circ}$22$'$59\farcs0, J2000.0) (Herbst \& Racine 1976), S156 (23$^h$05$^m$24$^s$.80, $60^{\circ}$08$'$14\farcs0) (Hoglund \& Gordon 1973), DR17/L906 (20$^h$35$^m$06$^s$.16, $42^{\circ}$20$'$23\farcs7) (Schneider et al. 2006) and M17/M18 (18$^h$20$^m$47$^s$.11, $-16^{\circ}$10$'$17\farcs5) (Downes et al. 1980).  "..." indicates that the corresponding spectra could not be detected or data were available but no reliable fit could be made. H$_2$CO data were selected from Tang et al. (2013). Values marked with "Y" are plotted as squares in Fig.3 and the data have been converted from the antenna temperature to the brightness temperature of the line.}

\begin{figure}[h]
\vspace*{2mm}
\begin{center}
\includegraphics[width=15cm]{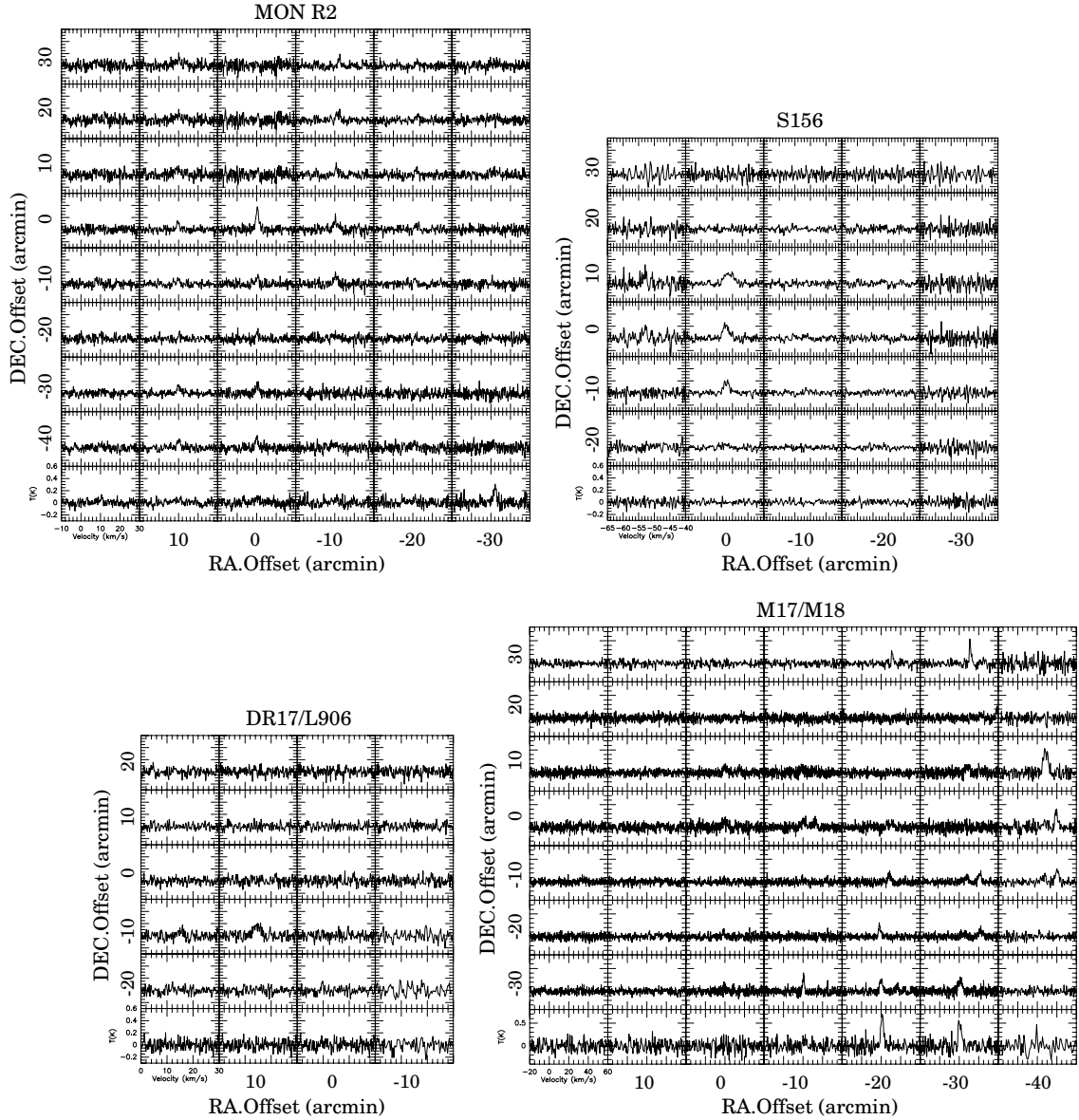}
\end{center}
\caption{The spectra of C$^{18}$O (1--0) lines toward MON R2, S156, DR17/L906, and M17/M18 regions.}
\end{figure}

\label{lastpage}

\end{document}